%% file: main.tex
\documentclass[final,authoryear,3p,times,twocolumn,fleqn]{elsarticle} 

\usepackage{filecontents}
\usepackage{textcomp}

\usepackage{amssymb}
\usepackage{amsmath}

\usepackage{lineno}

\journal{Earth and Planetary Science Letters}

\begin{document}

\begin{frontmatter}


\title{Inefficient volatile loss from the Moon-forming disk: reconciling the giant impact hypothesis and a wet Moon}

\author{Miki Nakajima$^{1,2}$}
\ead{mnakajima@carnegiescience.edu}
\author{David J. Stevenson$^2$}
\address{$^1$Department of Terrestrial Magnetism, Carnegie Institution for Science, 5241 Broad Branch Rd NW, Washington, DC 20015, USA. \\
$^2$Division of Geological and Planetary Sciences, California Institute of Technology, 1200 E California Blvd, MC 150-21, Pasadena, CA 91125, USA.}

\begin{abstract}
\input{abstract}

\end{abstract}
\begin{keyword}
Moon; giant impact; volatiles; lunar water; volatile loss; hydrodynamic escape

https://doi.org/10.1016/j.epsl.2018.01.026


%
\end{keyword}
\end{frontmatter}
%
\section{Introduction}
\label{intro}
\input{introduction}

%
\section{Model}
\label{model}
\input{model}

\section{Results}
\label{results}

\input{results}

\input{figures}
\section{Discussion}
\label{discussions}
\input{discussions}

\section{Conclusions}
\label{conclusion}
\input{conclusion}

%
%
%
\section*{Acknowledgement}
\label{acknowledgement}
\input{acknowledgement}

\setcounter{section}{1}
\section*{Supplementary Information}
\label{Appendix}
\input{appendix}

\bibliographystyle{model2-names}
\bibliography{bib_moonwater.bib}







\end{document}

%% file: abstract.tex
The Earth's Moon is thought to have formed from a circumterrestrial disk generated by a giant impact between the proto-Earth and an impactor approximately 4.5 billion years ago.
Since this impact was energetic, the disk would have been hot (4000-6000 K) and partially vaporized (20-100 \% by mass).
This formation process is thought to be responsible for the geochemical observation that the Moon is depleted in volatiles (e.g., K and Na). 
To explain this volatile depletion, some studies suggest the Moon-forming disk was rich in hydrogen, which was dissociated from water, and it escaped from the disk as a hydrodynamic wind accompanying heavier volatiles (hydrodynamic escape). 
This model predicts that the Moon should be significantly depleted in water, but this appears to contradict some of the recently measured lunar water abundances and D/H ratios that suggest that the Moon is more water-rich than previously thought.
Alternatively, the Moon could have retained its water if the upper parts (low pressure regions) of the disk were dominated by heavier species because hydrogen would have had to diffuse out from the heavy-element rich disk, and therefore the escape rate would have been limited by this slow diffusion process (diffusion-limited escape). 
To identify which escape the disk would have experienced and to quantify volatiles loss from the disk, we compute the thermal structure of the Moon-forming disk considering various bulk water abundances (100-1000 ppm) and mid-plane disk temperatures (2500-4000 K). Assuming that the disk consists of silicate (SiO$_2$ or Mg$_2$SiO$_4$) and water and that the disk is in the chemical equilibrium, our calculations show that the upper parts of the Moon-forming disk are dominated by heavy atoms or molecules (SiO and O at $T_{\rm mid} > 2500 - 2800$ K and H$_2$O at $T_{\rm mid} < 2500 - 2800$ K) and hydrogen is a minor species. This indicates that hydrogen escape would have been diffusion-limited, and therefore the amount of lost water and hydrogen would have been small compared to the initial abundance assumed. 
This result indicates that the giant impact hypothesis can be consistent with the water-rich Moon.
Furthermore, since the hydrogen wind would have been weak, the other volatiles would not have escaped either.
Thus, the observed volatile depletion of the Moon requires another mechanism.

%% file: introduction.tex
It is widely accepted that the Earth's Moon formed by a collision between the proto-Earth and an impactor approximately 4.5 billion years ago \citep{HartmannDavis1975, CameronWard1976}. This impact created a partially vaporized disk around the planet, from which the Moon accreted. 
In the standard version of this hypothesis, the impactor was approximately Mars-sized and the impact velocity was close to the escape velocity \citep{CanupAsphaug2001}.
This model has been favored because it can explain several observed aspects of the Earth-Moon system, such as its angular momentum, the lunar mass, and the small iron core of the Moon. 
However, the model cannot easily explain the observation that the Earth and Moon have identical or strikingly similar isotopic ratios (e.g., oxygen and tungsten, \citealt{Wiecheretal2001, Herwartzetal2014, Youngetal2016, Kruijeretal2015, Toubouletal2015}) given that the impact simulations indicate that most of the disk materials originate from the impactor, which presumably had different isotopic ratios from the Earth.
It may be, however, possible that the impactor happened to have similar isotopic ratios to those of Earth because the inner solar system may have been well-mixed \citep{Dauphas2017}. This idea might bolster the standard model, but it still requires an explanation for the identical tungsten isotopic ratios. 

Alternatively, \cite{CukStewart2012} suggest that a small impactor hit a rapidly rotating Earth while \cite{Canup2012} suggests that two half Earth-sized objects collided. 
In these models, the composition of the disk is similar to that of the Earth, and therefore, the isotopic similarities could be naturally explained. 
These new models are promising alternatives, but they may predict that the Earth's mantle becomes mixed by the energetic impact \citep{NakajimaStevenson2015}. This appears to contradict observed anomalies of short-lived isotopes indicating that the Earth has never been completely mixed (e.g., $^{182}$Hf-$^{182}$W, \citealt{Willboldetal2011, Toubouletal2012, Rizoetal2016, Mundletal2017} and noble gases, \citealt{Mukhopadhyay2012}). A recent suggestion that the Moon could have formed as the outcome of merging of smaller Moons and multiple impacts \citep{Rufuetal2017} revives an old idea requiring some specific, perhaps unlikely dynamical conditions to be acceptable.

In addition to the isotopic ratios of the Earth-Moon system, the chemical compositions provide further essential information.
The giant impact has been thought to be at least partly responsible for the observation that the Moon is depleted in volatiles, such as K, Rb, Na, and other volatile elements \citep[e.g.,][]{Krahenbuhletal1973b, Ringwoodetal1987}. 
Smaller K/Th and K/U ratios of the Moon than those of the Earth also indicate that the Moon is depleted in volatiles (K is more volatile than Th and U) \citep{Teraetal1974}.
The Moon-forming disk was hot and partially vaporized (up to 4000-5000 K and 20-30\% for the standard case and 6000-7000 K and 80-90\% for the recent models, \citealt{NakajimaStevenson2014}). 
Hydrogen, which would have been dissociated from water at this high temperature, may have escaped from the hot Moon-forming disk as a wind (hydrodynamic escape) together with heavier atoms and molecules \citep{GendaAbe2003, DeschTaylor2013}.

This model predicts that the Moon also lost a significant amount of water, but this appears to be inconsistent with some of the measured lunar water abundances.
Determining the bulk lunar water abundance is an active area of research; 
based on these measurements and modeling of the lunar interior evolution \citep[e.g.,][]{Boyceetal2010, McCubbinetal2010, Haurietal2011, Huietal2013, Saaletal2013}, the bulk water content of the Moon has been estimated to range from $<10$ ppm \citep{ElkinsTantonGrove2011} to a few hundred ppm \citep{Huietal2013, Haurietal2011, Haurietal2015, MillikenLi2017}. 
\cite{Linetal2017} suggest that the crustal thickness estimated by GRAIL (34-43 km, \citealt{Wieczoreketal2013}) can be consistent with an initially deep ($\sim 700$ km) lunar magma ocean with presence of water (270 - 1650 ppm).
On the other hand,  work using Cl and F in addition to H in apatite suggests that the water content of the Moon can be much lower \citep[e.g.,][]{Boyceetal2014}.
These results cover a wide range: the Moon could be ``wet'', which indicates here that the Moon is as water-rich as Earth (Earth's bulk water abundance is estimated as a couple of hundred ppm, \citealt{McDonoughSun1995}) or could be drier ($<$ 100 ppm).
Needless to say, the possibility that water is heterogeneously distributed within the Moon \citep{RobinsonTaylor2014} makes it even more difficult to estimate the bulk lunar water abundance based on a small set of samples.

If a significant amount of water escaped from the disk, the lunar D/H ratios should be more enhanced than that of the Earth because H is lighter and would have escaped more efficiently than D. However, analyses of pristine lunar water suggest that lunar D/H ratios may be comparable to the terrestrial values, which may indicate that water loss was insignificant  \citep[e.g.,][]{Saaletal2013}.
It should be noted that measuring the bulk content of the indigenous water and D/H ratio is a very challenging task because the available lunar samples are limited and because a number of processes, including fractional crystallization, degassing, solar wind irradiation, and cosmic-ray spallation, would likely alter the original values.

Thus, the Moon is depleted in some volatiles, but it may or may not be depleted in water. 
To understand the history of the lunar volatiles, we propose to reevaluate the water loss mechanism. \cite{DeschTaylor2013} suggest that hydrodynamic escape could have occurred and blew off the disk atmosphere when the disk temperature is 2000 K and the mean molecular weight of the disk is $\bar{m}=6$ g mol$^{-1}$ (i.e., water in the disk was dissociated to 2H and O). Conventionally, hydrodynamic escape from a planetary atmosphere occurs when the Jeans parameter $\lambda \equiv GM_\oplus \bar{m}/RTr'$ is smaller than $\sim$ 2 \citep{Parker1963} (the exact number of this criterion can vary depending on the atmospheric composition, \citealt{Volkovetal2011}, and the geometry, \citealt{DeschTaylor2013}).
Here, $G$ is the gravitational constant, $M_\oplus$ is the Earth mass, $R$ is the gas constant, $T$ is the temperature, and $r'$ is the distance from the planet.
The work done by \cite{DeschTaylor2013} is certainly insightful, but an important aspect here is that this criterion of $\lambda$ has been developed for a gas that behaves as a material with a single molecular weight (for example the solar wind, which is primarily hydrogen), and it is necessary to understand if this model is applicable to the specific system of interest.
If the disk were dominated by heavier elements that were gravitationally bound (i.e., the escape fluxes of the heavy elements were negligible), the hydrodynamic escape model is no longer valid.
For hydrogen to escape from a disk dominated by heavier elements, it must diffuse out from the heavy elements that are gravitationally bound to the planet-disk system.
Thus, the hydrogen escape rate is limited by this diffusion process and this is called diffusion-limited escape, which is much slower than the ``blow off'' hydrodynamic escape. This type of hydrogen escape likely occurred from early planetary atmospheres \citep[e.g.,][]{Hunten1973, Zahnleetal1990}.

As an example, consider a disk that is dominated by a light element $i$ and heavy element $j$.
We assume here that $i$ and $j$ are hydrogen and oxygen, respectively, and that their mole fractions are $f_i=\frac{n_i}{(n_i+n_j)}=f_{\rm H}=0.1$, where $n_i$ and $n_j$ are the number densities of the element $i$ and $j$, respectively. 
The upper limit of the diffusion-limited escape rate is described as \citep{Hunten1973},
\begin{equation}
\phi_l =  b_{ij} f_i \left(\frac{1}{H_j} - \frac{1}{H_i} \right)\sim b_{ij} f_i/H_j,
\label{limitflux}
\end{equation} 
where $b_{ij}$ is the binary collision parameter between elements $i$ and $j$, $f_i$ is the mole fraction of the element $i$, $H_i(=RT/m_i g$, where $m_i$ is the molecular weight of the element $i$) is the scale height of the element $i$. The subscripts $i$ and $j$ represent the elements $i$ and $j$, respectively. $R$ is the gas constant, and $g$ is the gravity.  The last approximation is valid when $H_i \gg H_j$. Here, we are assuming that the heavy element $j$ is not escaping from the system.

Under the hard-sphere approximation, $b_{ij}$ is described as \citep{ChamberlainHunten1987}, 
\begin{equation}
b_{ij}=\frac{3}{64Q}\left(2\pi RT\frac{m_i+m_j}{m_i m_j}\right)^{\frac{1}{2}},
\label{eq:b1}
\end{equation} 
where
\begin{equation}
Q=\frac{\pi}{16}(\sigma_i+\sigma_j)^2.
\label{eq:Q}
\end{equation} 
$\sigma$ is the collision diameter.
Assuming $i$ is atomic hydrogen and $j$ is atomic oxygen, the hydrogen escape flux becomes $1.83 \times10^{15}$ atoms m$^{-2}$s$^{-1}$ ($\sigma_i=2 \times53$ pm, $\sigma_j=2 \times60$ pm, $m_i$=1 g mol$^{-1}$, $m_j$=16 g mol$^{-1}$, $T=2000$ K, $f_i=0.1$, $r=3 R_\oplus$, $z=3 R_\oplus$, $r'=\sqrt{r^2 + z^2}=4.2 R_\oplus$ where $r$ is the horizontal distance from the planetary spin axis and $z$ is the vertical distance from the disk mid-plane, and $g=GM_\oplus z/r'^3$. The choices of these parameters are discussed in Section \ref{Hloss}). 
Assuming the surface area of the disk is $2\pi((5 R_\oplus)^2 - R_\oplus^2)$ and the disk life time is 1000 years, the total amount of lost hydrogen is $5.86\times 10^{14}$ kg and the equivalent amount of water is $5.86  \times10^{14} \times(18/2) = 5.27 \times10^{15}$ kg. It should be noted that this surface area is likely an upper limit because part the disk outside of the Roche radius would fragment.  
If the total mass of the disk is 1.5 lunar masses and the disk contains 100 ppm of water, then the mass fraction of the lost water with respect to the total water (i.e., the water loss mass fraction) is $4.78 \times 10^{-4}$. 
This is too small to have a significant effect on the interpretation of measurements of the water abundance or D/H ratio of the Moon.

Thus, determining the escape mechanism is highly important for estimating the volatile loss from the Moon-forming disk.
In this paper, we determine the structure of the disk and find that the upper parts of the disk are dominated by heavy atoms and molecules (Sections \ref{diss} and \ref{Hloss}).
This indicates that the hydrogen escape is likely diffusion-limited and that the amount of hydrogen and other volatiles escaping from the disk is too small to be observed.
This may indicate that the observed volatile loss (e.g., K, Na, and Rb) would require another explanation as discussed in Section \ref{volatiles}.

%% file: model.tex
We assume that the disk has a liquid layer in the mid-plane that is sandwiched by vapor layers (this picture is similar to Figure 3 in \citealt{PahlevanStevenson2007}, but we assume that the disk is isolated from the Earth's atmosphere).
We assume that the disk consists of water and silica (SiO$_2$ except Section \ref{mg2sio4}, where Mg$_2$SiO$_4$ is considered). 
To estimate the hydrogen abundance in the upper parts of the disk, we simply investigate the vertical disk structure at a certain radial location (at $r=3R_\oplus$ except Figure \ref{fig:mass_loss}) given a mid-plane temperature $T_{\rm mid}$ instead of modeling the whole disk structure. 
We make this simplification mainly because our outcome is insensitive to the detailed radial disk structure and partly because the time-dependent disk structure is not well known.
The effect of $r$ on the water loss is considered in Section \ref{Hloss}.

The surface density of the disk is assumed to be $5\times 10^7 $ kg m$^{-2}$ based on previous work \citep{Canupetal2013, NakajimaStevenson2014}.
We also test $10^8 $ kg m$^{-2}$ (not shown), but the outcome is similar to the case presented here. 
The model parameters are the bulk water content of the disk (100, 500, and 1000 ppm) and mid-plane temperature $T_{\rm mid}$ (2500 - 4000 K).
This range can be thought of as representing different stages in the cooling of the disk or different giant impact models (or both).

To estimate the amount of water loss, we first determine the mixing ratio of silicate vapor and water vapor in the upper parts of the disk without considering dissociation of molecules (Section \ref{vertical}).
Based on the pressure and temperature ranges obtained, we estimate the mole fractions of the molecules and atoms considering dissociation given that the system in the chemical equilibrium (Section \ref{thermal}). If the disk is dominated by heavy elements, hydrogen escape is diffusion-limited. The amount of lost hydrogen and water is estimated using Equation (\ref{limitflux}).

\subsection{Boundary condition at the liquid-vapor interface}
At the liquid-vapor interface (at the mid-plane), the partial pressure of water is given as \citep{AbeMatsui1986} 
\begin{equation}
p_{\rm H_2O}=\left(\frac{y_{\rm H_2O }(\rm{wt} \%)}{2.08\times 10^{-4}}\right)^\frac{1}{0.54} {\rm (Pa)}, 
\label{eq:PH2O}
\end{equation} 
where $y_{\rm H_2O}$ is the mass fraction of water in the liquid.
The saturation vapor pressure of pure SiO$_2$ liquid (2000-6000 K) is written as (the units are modified from \citealt{VisscherFegley2013})
\begin{equation}
p^*_{\rm SiO_2}=p_0 \exp(-L/RT), 
\label{eq:PSiO2}
\end{equation} 
where $p_0=1.596\times 10^{13}$ Pa and $L=4.96\times 10^5$ J mol$^{-1}$.
The total pressure at the interface becomes $p=(1-x^l_{\rm H_2O})p^*_{\rm SiO_2}+p_{\rm H_2O}$, where $x^l_{\rm H_2O}$ is the mole fraction of water in the liquid ($x^l_{\rm H_2O}=\frac{y_{\rm H_2O}}{18} ( \frac{y_{\rm H_2O}}{18}+\frac{100-y_{\rm H_2O}}{60})^{-1}$).

\subsection{Vertical structure of the disk}
\label{vertical}
We assume that the disk is in the radiative-convective equilibrium, which implies that lower part of the disk (i.e., small $z$) is convective and the upper part (i.e., large $z$) is radiative.  
In previous studies, the Moon-forming disk is assumed to be convective \citep{ThompsonStevenson1988, GendaAbe2003, Ward2012} or isothermal in the vertical direction \citep{CharnozMichaut2015}, but we modify this assumption in order to determine the temperature of the upper part of the disk. 
Provided the radial flow is highly subsonic, the disk is in hydrostatic equilibrium, i.e., $dp/dz=-\rho g$ where $p$ is the pressure, and $\rho$ is the density. The disk self-gravity is ignored.

The convective region of the disk follows the moist pseudoadiabat curve, which assumes that the partial pressure of silicate vapor is equal to its saturation vapor pressure.
The amount of water extracted by solution in the silicate rain is small because of the modest water vapor pressure. This implies that only silicate rains out and this rain-out is efficient (if it were not then our conclusion that the water escape is minor would be even stronger since the upper atmosphere would contain droplets or particles that impede outflow). 
A moist pseudoadiabatic lapse rate is described as \citep{Nakajimaetal1992},
\begin{equation}
\left(\frac{\partial T}{\partial p}\right)_s= \frac{\frac{RT}{pC_{p,{\rm H_2O}}}+\frac{x_s}{x_{\rm H_2O}}\frac{L}{pC_{p,{\rm H_2O}}}}{x_{\rm H_2O}+x_s\frac{C_{p,s}}{C_{p,{\rm H_2O}}}+\frac{x_s}{x_{\rm H_2O}}\frac{L^2}{RT^2 C_{p,{\rm H_2O}}}},
\label{eq:momentum}
\end{equation} 
where $x_s$ is the mole fraction of saturated silicate ($=p_{\rm SiO_2}^*(T)/p$) and $x_{\rm H_2O}(=1-x_s)$ is the mole fraction of water. The subscripts, s and H$_2$O represent the parameters of SiO$_2$ and water, respectively.
It should be noted that dissociation of SiO$_2$ and H$_2$O are not considered in this section and it would be considered only in Section \ref{thermal}.
The effect of this simplification on the temperature profile is further discussed in Section \ref{validation2}).
$C_p$ is the specific heat and $C_{p,{\rm s}}=62$ JK$^{-1}$mol$^{-1}$, and $C_{p, {\rm H_2O}}=55.7$ JK$^{-1}$mol$^{-1}$ (at 1 bar at 3000K, \citealt{Chaseetal1985}).

The optical depth throughout the disk is described as \citep{Nakajimaetal1992}
\begin{equation}
d\tau=(\kappa_s x_s m_s+\kappa_{\rm H_2O} x_{\rm H_2O} m_{\rm H_2O})\frac{dp}{m_{\rm ave}g},  
\label{eq:tau}
\end{equation} 
where $\tau$ is the optical depth, $\kappa$ is the absorption coefficient, $m$ is the molecular weight, and $m_{\rm ave}$ is the average molecular weight ($m_s=m_{\rm SiO_2}=60$ g mol$^{-1}$ and $m_{\rm H_2O}=18$ g mol$^{-1}$).
Here, $\kappa_s=0.1$ m$^2$kg$^{-1}$ \citep{ThompsonStevenson1988} and $\kappa_{\rm H_2O}=0.01$ m$^2$kg$^{-1}$ are used \citep{Nakajimaetal1992}.
We assume that the disk is treated as a gray atmosphere and the upward radiation flux at given $\tau$ is written as 
\begin{align}
F_\uparrow(\tau) &=\frac{3}{2}\int_\tau^{\tau_{\rm mid}} \pi B(\tau')\exp \left[-\frac{3}{2} (\tau'-\tau) \right]d\tau' \nonumber \\
& +\pi B(\tau_{\rm mid})\exp\left[\frac{3}{2} (\tau-\tau_{\rm mid})\right],
\label{eq:Fup}
\end{align} 
where $\tau_{\rm mid}$ is the optical depth at the mid-plane (the liquid-vapor boundary).
Likewise, the downward radiation flux is written as
\begin{equation}
F_\downarrow(\tau)=\frac{3}{2}\int_0^{\tau} \pi B(\tau')\exp \left[-\frac{3}{2} (\tau'-\tau) \right]d\tau' .
\label{eq:Fdown}
\end{equation} 
The net upward flux is written as $F(\tau)=F_\uparrow(\tau)-F_\downarrow(\tau)$. Here, $\pi B=\sigma T^4$ where $\sigma$ is the Stefan-Boltzmann constant.
In the radiative part of the disk, the temperature profile follows the relationship
\begin{equation}
\pi B=\sigma T^4(\tau)=\frac{1}{2}F_{\uparrow \rm top}(\frac{3}{2}\tau+1),
\label{eq:temp_rad}
\end{equation} 
where $F_{\uparrow \rm top}$ is the radiation flux from the top of the atmosphere (where $\tau=0$).
The transition (tropopause) between the convective lower disk and radiative upper disk is iteratively determined.
First, assuming that the whole atmosphere is convective, we compute the vertical structure of the disk until the condition described by Equation (\ref{eq:temp_rad}) is met.
This provides $\tau_{\rm tp}$, $T_{\rm tp}$, and $F_{\rm tp}$ where $F_{\rm tp}=F_{\uparrow \rm top}$.
We recalculate Equations (\ref{eq:Fup}) and (\ref{eq:Fdown}) and adjust the value of $F_{\uparrow \rm top}$ assuming that the lower parts of the disk are convective and the upper parts of the disk are radiative (generally speaking, this correction is very minor).

\subsection{Dissociation of molecules}
\label{thermal}
At high temperatures and low pressures, molecules in the disk can dissociate. 
For example, assuming that the system is in the chemical equilibrium, the dissociation of SiO$_2=$ SiO + O is described as
\begin{equation}
K_{\rm th}=\frac{p_{\rm SiO} p_{\rm O}}{p_{\rm SiO_2}}=\exp(-\Delta G^0 /RT),
\label{eq:K}
\end{equation} 
where $K_{\rm th}$ is the equilibrium constant, and $p_{\rm SiO}$, $p_{\rm O}$, and $p_{\rm SiO_2}$ are the partial pressures of SiO, O and SiO$_2$ in bar. $G^0$ is the Gibbs free energy under the standard conditions ($\Delta G^0=\Delta H^0 - T\Delta S^0$). 
Here, $\Delta H^0$ is the change in the Helmholtz energy and $\Delta S^0$ is the change in the entropy. 
We assume that $\Delta H^0$ and $\Delta S^0$ are not sensitive to temperature and pressure.
The rest of the reactions and the thermal constants are listed in Table \ref{thermK}.

\begin{center}
\begin{table*}[ht]
\scalebox{1.0}{
\hfill{}
\begin{tabular}{|c c c |}
\hline
Reaction & $\Delta S^o$ (J/mol K)& $\Delta H^o$ (kJ/mol)   \\ 
\hline
SiO$_2$  = SiO  + $\frac{1}{2}$O$_2$  &85.2  &205  \\
SiO  = Si  + $\frac{1}{2}$O$_2$  &85.1  &550 \\   
MgO  = Mg + $\frac{1}{2}$O$_2$   &38.0  &88 \\   
O$_2$  = 2O  &117  &498  \\ 
H$_2$O = H$_2$  + $\frac{1}{2}$O$_2$ &44.4  &242   \\
H$_2$ = 2H  &98.8  &436   \\

\hline
\end{tabular}}
\hfill{}
\caption{Reactions and thermodynamic constants \citep{Chaseetal1985}. All the elements are in the vapor phase. }
\label{thermK}
\end{table*}
\end{center}


\subsection{Homopause location}
\label{homoloc}
The escape flux of hydrogen is determined by thermal properties at the homopause, where eddy diffusion $K$ equals molecular diffusion $D$ (the molecular diffusion coefficient is described as $D=b_{ij}/N$ where $N$ is the number density). 
At lower parts of the disk (small $z$), eddy diffusion is more efficient than molecular diffusion ($K>D$), and therefore the disk is homogenized (i.e., the mole fractions of atoms and molecules are constant).
On the other hand, in the upper parts of the disk (large $z$), molecular diffusion becomes more dominant ($K<D$), and therefore each molecule or atom has its own scale height (i.e. light elements are more abundant at large $z$). 
If a hydrogen atom (or any light element) is present above the homopause, the atom can easily escape from the disk. 
However, this escape rate cannot exceed the hydrogen supply below the homopause. 
In other words, the supply rate is determined by how fast a hydrogen atom can diffuse from a heavy-element rich disk.
This is the definition of diffusion-limited escape as we briefly describe in Section \ref{intro}.
We estimate that the homopause pressure range is $\sim 10^{-4} - 10^{1}$ Pa, as we discuss in Section \ref{Appendix_homo_loc}. Even though the pressure range of homopause is not well defined, fortunately, our result is insensitive to this parameter.

%% file: results.tex
\subsection{Vertical structure of the disk}
\label{vertical_res}

The vertical structure of the disk is shown in Figure \ref{fig:vertical}. The left panels show the temperature-pressure structure and right panels show the mixing ratio of water ($x_{\rm H_2O}$). 
The top, middle, and bottom panels correspond to the cases when the bulk water abundance is  (a) 100 ppm, (b) 500 ppm, and (c) 1000 ppm.
The location of the homopause is indicated by the shade (a large uncertainty, as discussed in Section \ref{homoloc} and \ref{Appendix_homo_loc}).
The lower parts of the disk  (i.e., large $p$ and small $z$) are in the convection regime while the upper parts (i.e., small $p$ and small $z$) are in the radiative regime.
The location of this transition (i.e., tropopause) is indicated by the kink in the temperature profile.
This kink appears because the disk temperature continues to decrease in the convective regime as $z$ increases, whereas it is nearly constant in the radiative regime.

The water mixing ratio at the homopause is most sensitive to the mid-plane temperature.
As Figure \ref{fig:vertical} a2-c2 shows, at high mid-plane temperatures ($T_{\rm mid} >$ 2500 - 2800 K) the homopause is dominated by silicate vapor ($x_{\rm H_2O}<0.5$) because the partial pressure of the silicate vapor is larger than that of water vapor. In contrast, it is dominated by water vapor ($x_{\rm H_2O}>0.5$) at lower mid-plane temperatures ($T_{\rm mid} <$ 2500 - 2800 K) because most of the silicate condenses into liquid and the partial pressure of silicate vapor becomes negligible.
As the bulk water abundance increases, the water mixing ratio increases, but this is a relatively weak effect compared to the mid-plane temperature.

When the disk is dominated by silicate vapor, the temperature of the disk closely follows the saturation vapor pressure of SiO$_2$ (i.e., it is thermodynamically determined rather than determined by the heat flow). 
This is a universal curve (independent of disk mass, heat flow, and water provided $x_{\rm H_2O} \ll 1$). 
In contrast, when the disk is dominated by water ($x_{\rm H_2O} \sim 1$), the disk temperature profile is determined by the heat flow (i.e., the radiative boundary condition) and the temperature can be less than the value it would have if it were determined by vapor pressure equilibrium with silicate because there is no longer significant silicate present. The homopause height is $z \sim 3 R_\oplus$ for a silicate-rich disk, whereas $z \sim 5 R_\oplus$ for a water-rich disk.

\subsection{Atoms and molecules present in the disk}
\label{diss}
The atoms and molecules present at the homopause in the disk at a constant temperature are shown in Figures \ref{fig:diss_sio2} and \ref{fig:diss_h2o}.
In Figure \ref{fig:diss_sio2} (a) and (b), the water mixing ratio $x_{\rm H_2O}$ is 0.1 and 0.5, respectively while $x_{\rm H_2O}$ is 1 in Figure \ref{fig:diss_h2o}. 
Figure \ref{fig:diss_sio2} corresponds to cases where the mid-plane temperature is high (i.e. $T_{\rm mid}> 2500 - 2800$ K) and the disk is dominated by silicate vapor, while Figure \ref{fig:diss_h2o}  corresponds to the cases where the mid-plane temperature is relatively small (i.e. $T_{\rm mid}< 2500 - 2800$ K) and the homopause is dominated by water vapor.

In Figure \ref{fig:diss_sio2} (a), the main reactions are described as SiO$_2$ = SiO + O and H$_2$O = 2H + O at $p < 10^{-3}$ Pa.
Hydrogen mole fraction $f_{\rm H}$ at this pressure range is $\sim0.1$ (this is approximately estimated by  0.9 SiO$_2$ + 0.1 H$_2$O = 0.9 SiO + 1O + 0.2 H. $f_{\rm H}\sim 0.2/(0.9+1+0.2)=0.095$). 
A similar argument can be made for Figure \ref{fig:diss_sio2} (b), where 0.5 SiO$_2$ + 0.5 H$_2$O = 0.5 SiO + 1O + 1H, therefore $f_{\rm H} = 1/(0.5+2)=0.4$. 
This $f_{\rm H} $ is likely to be an overestimate because the homopause temperature is below 2000 K at $x_{H_2O}=0.5$ (for example, the homopause temperature is $1660$ K at $T_{\rm mid}=2500$ K and $p=10^{-2}$ Pa with 100 ppm of water as shown in a1, Figure \ref{fig:vertical}). $x_{\rm H_2O}>0.5$ can occur when the water abundance is 1000 ppm (c2, Figure \ref{fig:vertical} at $T_{\rm mid}=2800$ K), but 1000 ppm is likely to be larger than the actual bulk lunar water abundance (Section \ref{intro}).

When the homopause is dominated by water ($x_{\rm H_2O}\sim 1$), the homopause temperature is approximately 1600 K or below at $p < 1$ (Pa) (Figure \ref{fig:vertical}). 
Under these circumstances, H$_2$O exists as its molecular form, and thus, hydrogen H is not abundant due to this relatively low temperature as shown in Figure \ref{fig:diss_h2o} (a). If the homopause temperature is close to 2000 K (Figure \ref{fig:diss_h2o}b), H$_2$O is dissociated to 2H and O, and hydrodynamic escape would be expected from such a disk as suggested by \cite{DeschTaylor2013}. However, we suggest that this is unlikely to be the case for the lunar disk because the homopause temperature is smaller ($\sim 1600$ K).

\subsection{Water loss from the disk}
\label{Hloss}
Hydrogen loss from the disk is estimated from Equation (\ref{limitflux}).
First, we consider the case when the disk temperature is high and the disk is dominated by silicate vapor ($f_{\rm H} < 0.4$ as discussed in Section \ref{diss}). 
This is likely the case shortly after the Moon-forming impact when the disk temperature is close to the initial disk temperature (4000-6000 K, \citealt{NakajimaStevenson2014}). 
Assuming the disk is dominated by O with a minor amount of H ($f_{\rm H}=0.1, f_{\rm O}=0.9$), the lost water by diffusion-limited escape is estimated as 5.27 $\times10^{15}$ kg and the mass fraction of lost lunar water is $4.78 \times 10^{-4}$ as discussed in Section \ref{intro}. 

Now we consider the case when the homopause is dominated by water. Given the homopause temperature is 1600 K, $f_{\rm H}=0.3$, $m_2=18$ g mol$^{-1}$, $\sigma_2$ = $265$ pm, $z=5R_\oplus$ with bulk 100 ppm of water, the hydrogen escape flux is $1.64 \times 10^{15}$ atoms m$^{-2}$s$^{-1}$, the lost water mass is $4.73\times10^{15}$ kg and the mass fraction of lost water is $4.28\times 10^{-4}$. 

The mass fraction of water loss with various parameters ($T=1600$ K and 2000 K, $f_{\rm H}=0.1, 0.3,$ and 0.5, $r=3-7 R_\oplus$) is shown in Figure \ref{fig:mass_loss}, which confirms that water loss from the disk is minor ($<$ a few $10^{-3}$ under these conditions). The range of $r$ is taken from previous work ($r\leq 5-7 R_\oplus$, \citealt{Canupetal2013, NakajimaStevenson2014}).

%% file: figures.tex
\begin{figure*}
  \begin{center}
    \includegraphics[scale=1.0]{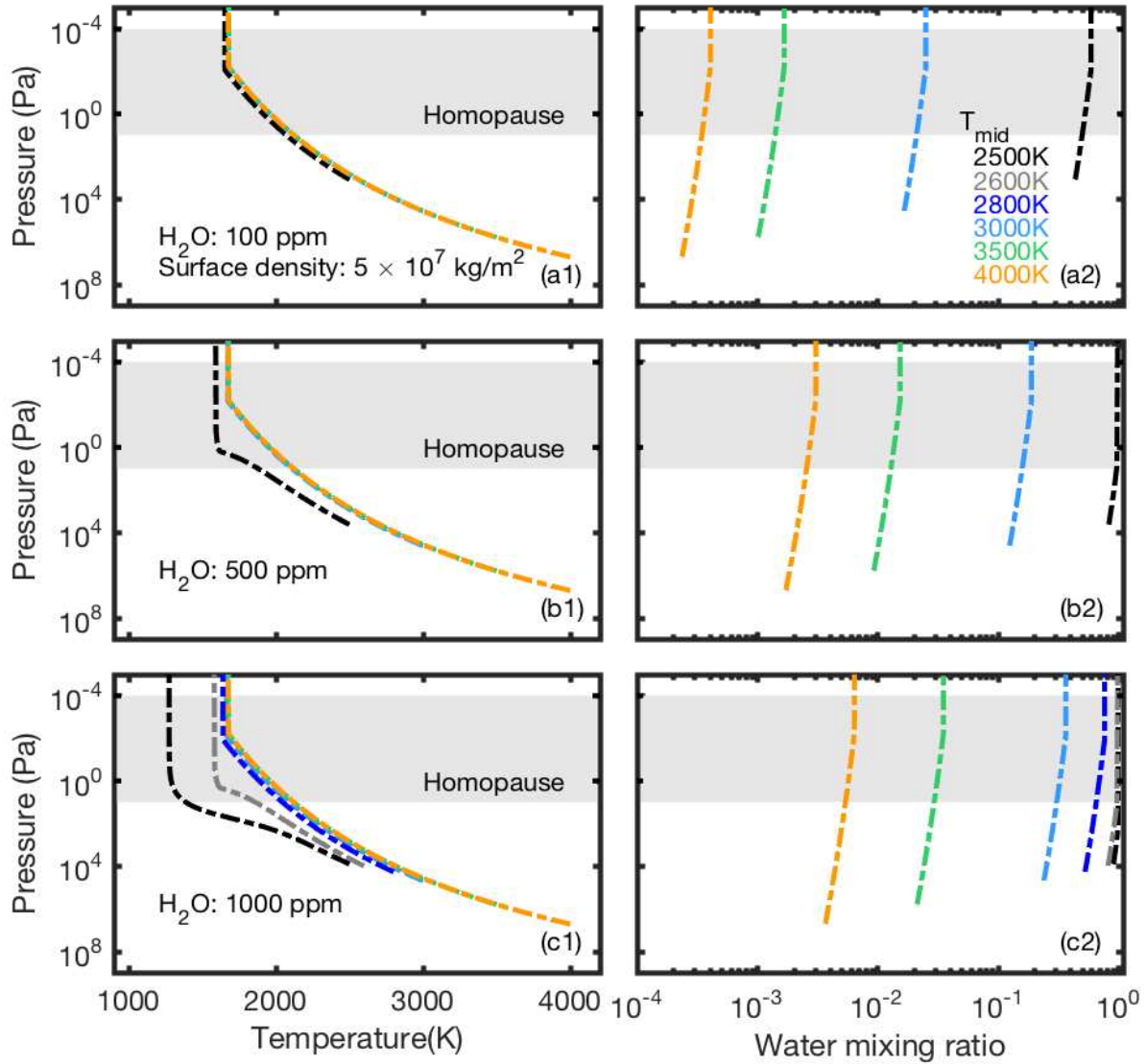}
  \end{center}
  \caption{Vertical structure of the Moon-forming disk with the surface density of $5\times 10^7$ kg m$^{-2}$. 
  The top, middle, and bottom panels represent cases when the bulk water abundance is 100 ppm, 500 ppm, and 1000 ppm, respectively.
The black, gray, blue, sky-blue, green, and orange lines correspond to $T_{\rm mid}=$2500 K, 2600 K, 2800 K, 3000 K, 3500 K, and 4000 K (2600 K and 2800 K are shown only for the case with 1000 ppm water). The homopause location is indicated by the shade. The left panels show the temperature whereas the right panels show the water mixing ratio ($x_{\rm H_2O}$) as a function of pressure. When the disk temperature is high and the bulk water abundance is low, the water mixing ratio is small and the pressure closely follows the saturation vapor pressure of SiO$_2$. In contrast, when the disk temperature is small, the upper disk atmosphere is dominated by water and the pressure is smaller than the saturated vapor pressure of SiO$_2$.
}
\label{fig:vertical}
\end{figure*}
%

%
\begin{figure*}
  \begin{center}
    \includegraphics[scale=0.8]{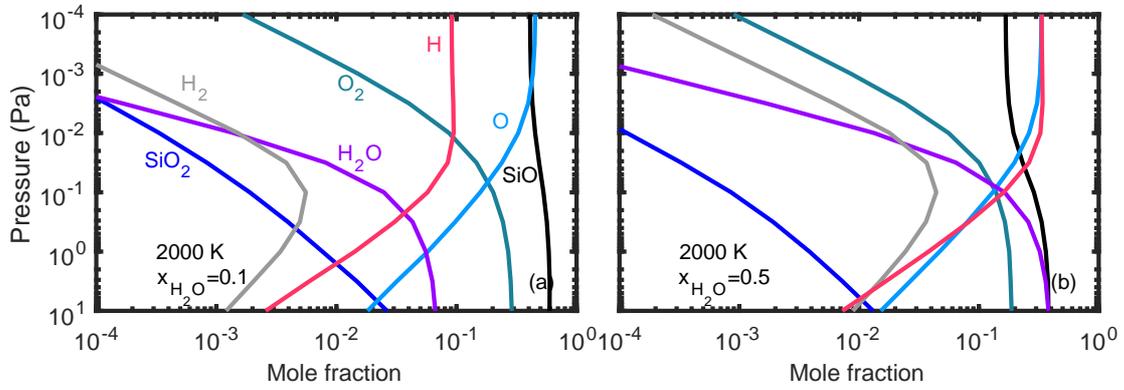}
  \end{center}
  \caption{Species that are present at the homopause at 2000 K. The y-axis corresponds to the homopause pressure (Section \ref{homoloc}). SiO$_2$, SiO, O$_2$, O, H$_2$O, H$_2$, and H are shown in blue, black, green, sky blue, purple, gray, and magenta (the Si mole fraction is too small to be shown in this figure). The left panel shows the dissociation of SiO$_2$ at $x_{\rm H_2O}=0.1$ and the right panel shows the case at $x_{\rm H_2O}=0.5$.
  In both cases, the hydrogen mole fraction $f_{\rm H}$ is small ($<$0.4).}
\label{fig:diss_sio2}
\end{figure*}
\begin{figure*}
  \begin{center}
    \includegraphics[scale=0.8]{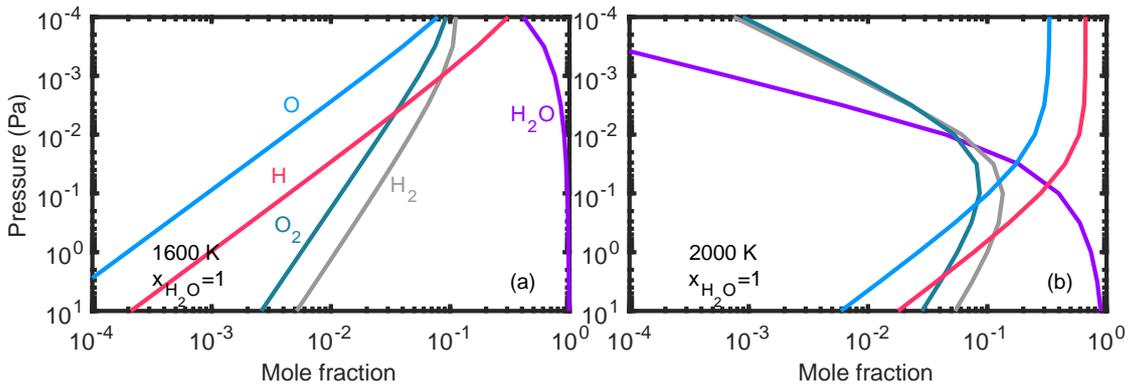}
  \end{center}
  \caption{Dissociation of pure H$_2$O at 1600 K (left) and 2000 K (right). At 1600K, H$_2$O is the dominant species, whereas H is dominant at low pressures at 2000 K. 
  The left panel (1600 K) better describes the Moon-forming disk because the homopause temperature is close to 1600K when the homopause is dominated by water.}
\label{fig:diss_h2o}
\end{figure*}
\begin{figure*}
  \begin{center}
    \includegraphics[scale=0.8]{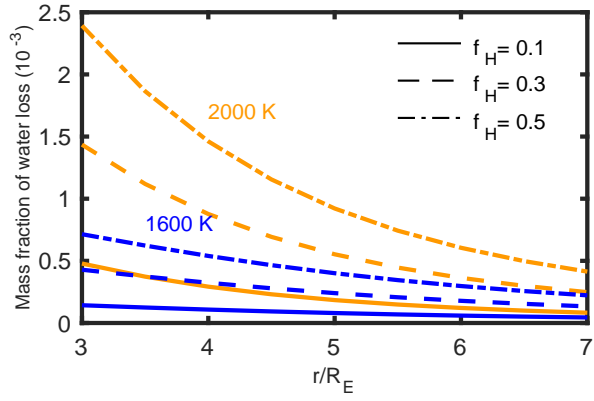}
  \end{center}
  \caption{Mass fraction of water loss with the various disk temperatures and the hydrogen mole fractions. The solid line, dashed line, and dotdash lines correspond to $f_{\rm H}=0.1, 0.3,$ and 0.5. The blue and orange lines represent 1600 K and 2000 K, respectively. For this calculation, we assume that the disk is dominated by H$_2$O at 1600K and by O at 2000 K, as discussed in Sections \ref{intro} and \ref{Hloss}.}
\label{fig:mass_loss}
\end{figure*}
\begin{figure*}
  \begin{center}
    \includegraphics[scale=0.8]{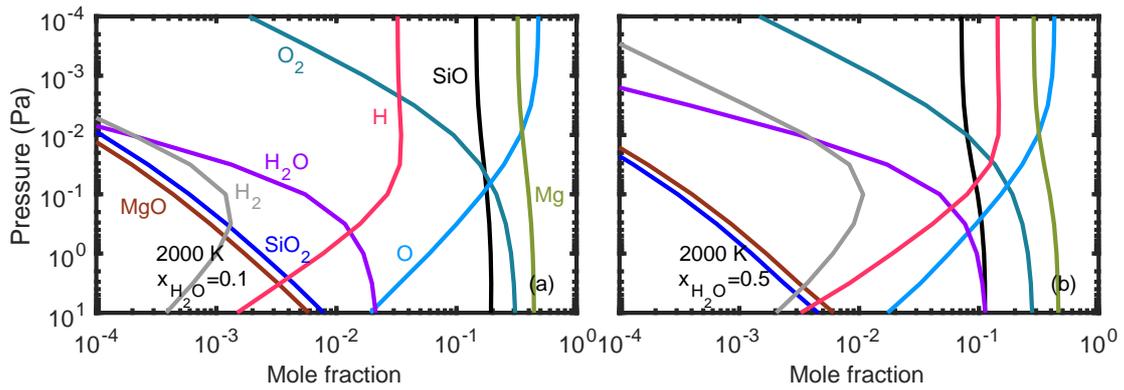}
  \end{center}
  \caption{Dissociation of Mg$_2$SiO$_4$ and H$_2$O. The color scheme is the same as Figure \ref{fig:diss_sio2} with additional elements (MgO and Mg are shown in brown and green). The hydrogen mole fraction is smaller than the case of SiO$_2$ because additional species (mainly Mg and O) are produced.}
\label{fig:diss_mg2sio4}
\end{figure*}

%% file: discussions.tex
\subsection{Mg$_2$SiO$_4$ disk}
\label{mg2sio4}
In Section \ref{results}, we only consider a disk that contains SiO$_2$ and H$_2$O, but the disk composition would be better modeled by Mg$_2$SiO$_4$. 
The disk temperature-pressure profile would not be significantly different given that the saturation vapor pressure of the bulk silicate Earth composition is similar to that of SiO$_2$ (Figure 3,  \citealt{VisscherFegley2013}), whereas species in the disk would be different. 
Figure \ref{fig:diss_mg2sio4} shows the mole fractions of the species present in a disk of Mg$_2$SiO$_4$ and H$_2$O.
At $x_{\rm H_2O}$ = 0.5, the main reactions are described as 0.5 Mg$_2$SiO$_4$=0.5 (2Mg + SiO + 3O) and 0.5 H$_2$O= 0.5 (2 H + O), which leads to $f_{\rm H}=0.22$. 
The presence of Mg decreases $f_{\rm H}$ and makes it even more difficult to lose hydrogen from the disk. 
$f_{\rm H}$ may decrease even further when more species (such as Al$_2$O$_3$ and CaO) that are expected to be in the disk are considered \citep[e.g.,][]{VisscherFegley2013, Itoetal2015}. We do not consider effects of Mg on the solubility of water in the magma even though it could dissolve as Mg(OH)$_2$ (for detailed discussion, see \citealt{Fegleyetal2016}).
Furthermore, some OH can be present in the disk (\citealt{Pahlevanetal2016} and personal communications with J. Melosh), which is not considered in this paper. It is possible that presence of OH can decrease $f_{\rm H}$ and the amount of water loss.

\subsection{Escape of other volatiles}
\label{volatiles}
In the previous sections, we only consider the escape of hydrogen, but other volatile elements, such as K and Na, can be considered in a similar framework.
\cite{Zahnleetal1990} consider the escape of a minor species with the presence of two major constituents.
The escape flux of a minor constituent $k$, $\phi_k$, is approximately described as,
\begin{equation}
\phi_k=F_k\phi_{\rm H} \left(\frac{1-\frac{m_k-m_{\rm H}}{m_2-m_{\rm H}} \frac{b_{{\rm H}k}}{b_{{\rm H}2}}+\frac{m_2-m_k}{m_2-m_{\rm H}}\frac{b_{{\rm H}k}}{b_{{\rm H}2}}F_2}{1+\frac{b_{{\rm H}k}}{b_{2k}}F_2}  \right),
\label{eq:zahnle_main}
\end{equation} 
where the subscripts H, 2, and $k$ represent hydrogen, heavy element that does not escape (corresponding to species such as SiO and O in the Moon-forming disk), and the element $k$, respectively.  
Here, $F_k=\frac{n_k}{n_H}$ and $F_2=\frac{n_2}{n_H}$ (the definition of $F_k$ is slightly different from $f_k=\frac{n_k}{n_H+n_i+n_k}$ as described in Section \ref{intro}).

Assumptions of this model are (1) $F_k  \ll F_2, 1$, and (2) $F_2$ is small. 
If the element $k$ is potassium and the element 2 is atomic oxygen, given that $\sigma_2=2\times 60$ pm, $\sigma_k=2\times 220$ pm, $m_2=16$ g mol$^{-1}$, $m_k=39.1$ g mol$^{-1}$, and $F_2=1$, and the equation above becomes $\phi_k=0.0683 F_k \phi_{\rm H}$. Assuming that $F_{\rm k} =  1$, and $\phi_H\sim 2 \times 10^{15}$ atoms m$^{-2}$s$^{-1}$, $\phi_k=1.37 \times 10^{14}$ atoms m$^{-2}$s$^{-1}$, which corresponds to $1.73 \times10^{15}$ kg. This is much smaller than the total potassium mass of the Moon is 2.64$\times 10^{19}$ kg if the potassium abundance of the Moon-forming disk (1.5 lunar masses) is the same as terrestrial values  ($240$ ppm, \citealt{McDonoughSun1995}).
Equation (\ref{eq:zahnle_main}) may not necessarily be useful for estimating other volatiles such as Na, because Na may be a dominant constituent instead of a minor constituent \citep{VisscherFegley2013}. However, it is still unlikely that this weak hydrogen flux would drag other heavier volatile elements.

Given that diffusion-limited escape is not an efficient mechanism for removing water or volatiles from the Moon-forming disk, the next question becomes how the Moon lost its volatiles or failed to efficiently accrete the volatiles.
A potential explanation is to accrete these volatiles onto Earth. 
\cite{Canupetal2015} suggest that some volatiles that were initially present in the disk could have been preferentially accreted onto the Earth at the end of the Moon accretion process. 
The amount is not a significant part of the total volatile budget of Earth and may only have a small effect on the isotopic ratios for the Moon because high disk temperature may not cause significant isotopic fractionation especially if it is described as equilibrium isotope fractionation.
Slightly higher lunar potassium isotopic ratios than those of the Earth may also indicate that the Moon forms from liquid part of the disk \citep{WangJacobsen2016, Locketal2016}.
An alternative is to lose volatiles directly from the Moon as or after it accretes (Section \ref{after}).

\subsection{Model validation}
\label{validation1}
First, in Section \ref{results}, we assume that heavy elements, such as oxygen O, would not escape from the disk, but this is an approximation. Some oxygen would escape, but the amount is likely to be limited. The escape regime of oxygen would be diffusion-limited because O would be too heavy to hydrodynamically escape. Consider a disk that consists of SiO and O, and $f_{\rm O}=0.5$, $T=2000$ K, $\sigma_1=2\times 60$ pm, $\sigma_2=320$ pm, $z=3R_\oplus$ then the escape flux of O becomes $1.88\times 10^{15}$ molecules m$^{-2}$ s$^{-1}$. This is too small to affect the oxygen abundance in the Moon. 

For the two cases discussed in Section \ref{Hloss} (silicate-rich and water-rich disks), the Jeans parameter $\lambda=GM_\oplus \bar{m}/RTr'$ is estimated as 26.9 ($\bar{m}=30$ g mol$^{-1}$, $T=2000$ K, and $r'=4.2R_\oplus$) and 14.6 ($\bar{m}=18$ g mol$^{-1}$, $T=1600$ K, and $r'=5.8R_\oplus$), both of which are much smaller than 2, suggesting that hydrogen escape is diffusion-limited. It should be noted, however, that the criterion (i.e., hydrodynamic escape at $\lambda <2$) is developed for an isotropic atmosphere with a single component, and therefore it may not be directly applicable to a Moon-forming disk which is neither isotropic nor a single component. 
Nevertheless, $\lambda$ being much larger than 2 further supports our model that the escape is diffusion-limited.

Furthermore, we use the formula of diffusion-limited escape developed for spherical geometry, but this geometry is not exactly applicable to the Moon-forming disk.
Nevertheless, we argue that the geometry becomes similar to a sphere given that the disk is extended in the vertical direction ($z\sim 3 - 5 R_\oplus$).
Furthermore, the physics of diffusion-limited escape is not sensitive to the geometry.  
We have not explicitly modeled the non-thermal aspect of the disk evolution, such as the decrease in surface density as disk material spreads outward as well as disk materials accreting to the Moon and Earth simultaneously.
However, much of this effect is simply equivalent to changing surface density and the results are not sensitive to this. Of course, aggregating material at large radius or onto Earth can only decrease the escape. 
Moreover, here we assume that the disk is isolated, but if the interaction of the disk and atmosphere is considered, the escape may become even smaller because volatiles in the disk may be replenished from Earth.

In this paper, we focus on diffusion-limited escape, but volatile escape can be limited by other sources. 
Some of the previous studies investigate energy-limited escape, where the escape is caused by energy input from the stellar EUV, and therefore the rate is limited by the stellar flux.
In that case, the escape rate is written as $\phi_{\rm EL}=\frac{\varepsilon F^* r'}{GMm^{'}}$, where $\varepsilon$ is the efficiency factor, $m^{'}$ is the averaged molecular weight, $F^*$ is the globally averaged EUV flux and Lyman $\alpha$ flux, 
which could have been  $10^{-2}-10^{-1}$ W m$^{-2}$ during this time period \citep[e.g.,][]{Zahnleetal1990, Baraffeetal2004, GendaIkoma2008}. At $\varepsilon = 0.3$ \citep{MurrayClayetal2009}, $m^{'}$=12.9 g mol$^{-1}$ (mixture of 0.3 H and 0.7 H$_2$O), $F^* = 0.1$ Wm$^{-2}$, $r'=4.2 R_\oplus$, this becomes $9.41 \times 10^{16}$ atoms m$^{-2}$s$^{-1}$ assuming the same disk surface area and disk life time discussed in Section \ref{intro}. This value is much larger than the diffusion-limited case.
However, this is equivalent to losing $4.61 \times10^{17}$ kg of water, which is still relatively small compared to the lunar bulk water ($1.10 \times 10^{19}$ kg at 100 ppm).  
This energy-limited escape can be more efficient if the escape occurs at higher $z$, where $f_{\rm H}$ is larger and $m^{'}$ is smaller, than the homopause, but then the rate would be ultimately limited by the hydrogen supply from below, and therefore the escape rate may not exceed the diffusion-limited escape rate. Thus, it is possible that energy-limited escape would contribute to water loss, but the extent is likely to be limited.
We further discuss model assumptions in Section \ref{validation2}.



\subsection{Comparison among different impact models}
Our calculations show that hydrogen escape is minor at $T_{\rm mid}=2500-4000$ K, where 4000 K is the initial disk temperature estimated for the canonical Moon-forming impact.
$T_{\rm mid}$ can be higher than 4000 K in the fast-spinning Earth and half-Earths models ($T_{\rm mid}=6000-7000$ K, \citealt{NakajimaStevenson2014}).
Nevertheless, even under this temperature rage, escape is still inefficient because $f_{\rm H}$ at the homopause is small.
Thus, water escape would be inefficient even outside of the temperature range considered in this paper.

\subsection{Lunar volatiles after the Moon formation}
\label{after}
In this paper, we focus on volatile loss during the disk phase, and it is possible that some volatiles were lost or added before and after this phase.
For example, the giant impact may have induced a vapor jet \citep{MeloshSonnet1986, Karato2014}, which may have removed some volatiles from the Moon-forming disk. 
This extent is difficult to quantify with the conventional impact method called smoothed particle hydrodynamics (SPH), where a fluid is expressed as a collection of spherical particles.
This is because the current SPH method cannot treat physics of two-phase (liquid-vapor) flow. 
Furthermore, some volatiles may have been added to or removed from the Moon before the lunar crust formation \citep{Bottkeetal2010, ElkinsTantonetal2011, ElkinsTantonGrove2011, Sharpetal2013, Haurietal2015, Haurietal2017}, which depends on the impact flux and crust formation time scale ($\sim 10 - 10^2 $ Myrs, \citealt{Meyeretal2010, ElkinsTantonetal2011}). 
We will further investigate evolution of lunar volatiles in the future.

%% file: conclusion.tex
We estimate the upper limit for hydrogen and volatile loss by thermal escape from the Moon-forming disk under various disk mid-plane temperatures (2500 - 4000 K) and the bulk water abundances (100, 500 and 1000 ppm). 
When the mid-plane disk temperature is large ($> 2500-2800$ K), the disk is dominated by silicate vapor. The major species in the upper part of the disk are SiO and O and the hydrogen mole fraction is small. In contrast, under low disk mid-plane temperature ($< 2500-2800$ K), the upper part of the disk is dominated by water and its temperature is  $\sim 1600$ K or below. 
In this temperature range, water stays in its molecular form (H$_2$O) and the hydrogen mole fraction is small as well. Since hydrogen is not the major element and other heavy elements (such as O, SiO, SiO$_2$, and H$_2$O) are the dominant species, hydrogen would need to diffuse out from this heavy-element rich disk. This escape regime is called diffusion-limited escape and it is an inefficient escape process. We estimate the total mass of lost water and volatiles, such as potassium, and find that the escape is inefficient and that it would not remove water or other volatiles from the disk to the extent that the loss is measurable in lunar rock samples. 
To remove volatiles from the Moon-forming disk or from the Moon, another mechanism, such as losing volatiles from the disk to the Earth or degassing from the lunar surface, would be required.

%% file: acknowledgement.tex
This work is supported by NASA Headquarters under the NASA Earth and Space Science Fellowship Program Grant NNX14AP26H and the Carnegie DTM Postdoctoral Fellowship. We would like to thank Masahiro Ikoma, Francis Nimmo, Cheng Li, Jay Melosh, Steve Desch, Erik Hauri, Alycia Weinberger, Peng Ni, and anonymous reviewers for helpful discussions.

%% file: appendix.tex
\renewcommand{\thesection}{A\arabic{section}}

\subsection{Homopause location}
\label{Appendix_homo_loc}
The location of the homopause is defined where $K=D$. The value of $K$ is highly uncertain because it is determined by dynamical processes that are imperfectly understood, such as upward propagating waves that break or other fluid dynamical instabilities (for example, magnetorotational instabilities, MRI, in the Moon-forming disk have been previously discussed, \citealt{CharnozMichaut2015, Carballidoetal2016, gammieetal2016}). 
An upper bound in the convective region follows from consideration of heat flow (essentially the bound provided by convective vigor). For a characteristic fluid velocity $v_{\rm conv}$ of $(F_{\rm conv}/\rho)^{\frac{1}{3}} \sim 10^2$ m s$^{-1}$, where $F_{\rm conv}$ is the convective flux and $\rho$ is the density, and the scale height of the disk $\bar{H} $ of hundreds of km, we expect $K\sim v_{\rm conv}\bar{H} \sim 10^7$ m$^2$ s$^{-1}$, but values in upper stably stratified regions can easily be many orders of magnitude lower. Simultaneously, the values can also be higher since the density is much lower (meaning that less energy per unit volume is required for mixing). 
The values of $K$ for planetary atmospheres are typically 10$^2$ - 10$^6$ m$^2$ s$^{-1}$ \citep{Atreya1986, Mosesetal2000, dePaterLissauer2010} but these are for much less energetic systems and for smaller characteristic length scales, and therefore a higher value is possible in our case. Fortunately, the value of $K$ does not matter much, as we now explain.

The upper limit of the escape flux for a light element $i$ (hydrogen), $\phi_i$ is approximately described as $\phi_i \leq   \phi_l \sim  b_{ij} f_i/H_j$ (Equation \ref{limitflux}) as discussed in Section \ref{intro}.
Importantly, the parameters in this equation change rather little even as the pressure and number density at the homopause change by many orders of magnitude. It should be noted that Equation (\ref{limitflux}) has a simple physical interpretation. Except for factors of order unity, it indicates that the escape flux is bounded above by $\sim nc$, where $n$ is the number density of the atom or molecule in question and $c$ is the sound speed for that species, with the number density being evaluated at the place where the mean free path for the dominant species is of order $H$ (this is the so-called exobase).

We can approximately estimate the homopause location based on the following two scenarios. If the region of the disk is dominated by oxygen atoms and has a small fraction of hydrogen atoms at 2000K, then $D\sim K$ at $p\sim 10^{-3}-10^{1}$ Pa
($m_i=1$ g mol$^{-1}$, $m_j=16$ g mol$^{-1}$, $\sigma_i$ = $2 \times 53$ pm, and $\sigma_j$ = $2\times60$ pm). 
If the disk is dominated by SiO at 2000K, this condition is met at $p\sim 10^{-4}-10^{0}$ Pa ($m_i=1$ g mol$^{-1}$, $m_j=30$ g mol$^{-1}$, $\sigma_i$ = $2\times53$ pm, and $\sigma_j$ = $320$ pm). Thus, the homopause pressure range is estimated as $\sim 10^{-4} - 10^{1}$ Pa as discussed in Section \ref{homoloc}.

\subsection{Further model validation}
\label{validation2}
There are further assumptions in the model in addition to the points raised in Section \ref{validation1}.
One of the underlying assumptions of hydrodynamic escape is that once the flow reaches its sound speed as it expands outwards and eventually escapes from the disk and Earth.
According to previous studies \citep{Walker1982, Zahnleetal1990}, this approximation is likely to be valid if the exobase is above the critical point. 
The critical point is where the velocity reaches the sound velocity.
This indicates that the criterion is $l_c<H_{1c}$, where $l_c$ is the mean free path and $H_{1c}$ is the scale height of a light element and the subscript $c$ describes the critical point. 
This is rewritten as \citep{Zahnleetal1990}
\begin{equation}
\phi_{l} \geq \frac{GM_\oplus}{r'^2} \frac{m_1}{RT}\frac{4b_{11}}{\sqrt{\pi}}
\label{condition}
\end{equation} 
This becomes $5.69 \times 10^{15}$ atoms m$^{-2}$s$^{-1}$  ($r'=4.24 R_\oplus$, $T=2000$ K,  $m_1=1\times10^{-3}$ g mol$^{-1}$), which is comparable to the hydrogen escape flux we estimate.
It is possible that the escape flux becomes smaller than this value and thus the condition above may not be met.
In this case, the hydrogen flux is very weak and heavy elements would not be dragged to space \citep{Zahnleetal1990}.
In other words, the diffusion-limited hydrogen escape rate is considered the upper limit and this makes our argument even stronger that hydrogen and volatile loss from the disk are minor.

Another assumption is that we ignore effects of dissociation on the lapse rate for simplicity (Section \ref{thermal}). The reaction of SiO$_2=$ SiO + O is endothermic and the absolute value of the enthalpy is comparable to that of SiO$_2$ condensation. This indicates that this reaction would affect the lapse rate, but this would still not change the final outcome because our result that the water escape is not efficient holds under the wide temperature range (2500-4000 K).